# High-Temperature Effects for Transition State Calculations in Solids

Chengxuan Ke[1], Chenxi Nie[1], and Guangfu Luo[1,2*]

[1]Department of Materials Science and Engineering, Southern University of Science and Technology, Shenzhen 518055, China

[2]Guangdong Provincial Key Laboratory of Computational Science and Material Design, Southern University of Science and Technology, Shenzhen 518055, China

*E-mail: luogf@sustech.edu.cn

**Abstract**

Transition state calculation is a critical technique to understand and predict versatile dynamical phenomena in solids. However, the transition state results obtained at zero Kelvin are often utilized for prediction or interpretation of dynamical processes at high temperatures, and the error bars of such approximation are largely unknown. In this benchmark study, all the major temperature effects including lattice expansion, lattice vibration, electron excitation, and band-edge shift are evaluated with first-principles calculations for defects diffusion in solids. With inclusion of these temperature effects, the notable discrepancies between theoretical predictions at zero Kelvin and the experimental diffusivities at high temperatures are dramatically reduced. Particularly, we find that the lattice expansion and lattice vibration are dominant factors lowering the defect formation energies and hopping barriers at high temperatures, but the electron excitation exhibits minor effects. In sharp contrast to typical assumption, the attempt frequency with lattice expansion and vibration varies significantly with materials: several THz for aluminum bulk but surprisingly over 500 THz for 4H-SiC. For defects in semiconductors, the band-edge shift is also significant at high temperatures and plays a vital role in the defect diffusion. We expect this study would help accurately predict the dynamical processes at high temperatures.

## I. INTRODUCTION

Transition state calculation of dynamical processes at high temperatures are essential to a range of fields, such as turbine engines, metallurgical processing, oxide film growth, and solid oxide fuel cells. As summarized in Fig. 1, high temperatures can induce lattice expansion, lattice vibration, electron excitation, and band-edge shift. These effects are coupled and impact several aspects of a dynamical process, including hopping distance, defect concentration, attempt frequency, and activation barrier. Because of the complexity, most previous studies utilized transition state results at zero Kelvin to predict the processes occurring at high temperatures.[1-11] Although limited efforts have made to include certain temperature effects in transition state calculations,[12-16] none of them included all the temperature effects and the differences between metals and semiconductors remain largely unknown. For instance, only the contribution of lattice vibration was considered for the hopping barrier of metallic solute diffusion in silicon.[17] The contributions of lattice vibration to defect concentration and hopping barrier were considered for vacancy intermediated self-diffusion in lithium, but the lattice constants and attempt frequency were fixed to values at zero Kelvin.[18] Neugebauer *et al.* carefully examined the anharmonicity and vacancies near the melting point of aluminum, while the impacts on dynamics was unexplored.[19] Therefore, a comprehensive evaluation of all the temperature effects consistently in representative systems are valuable to clarify their roles in dynamical processes at high temperatures.

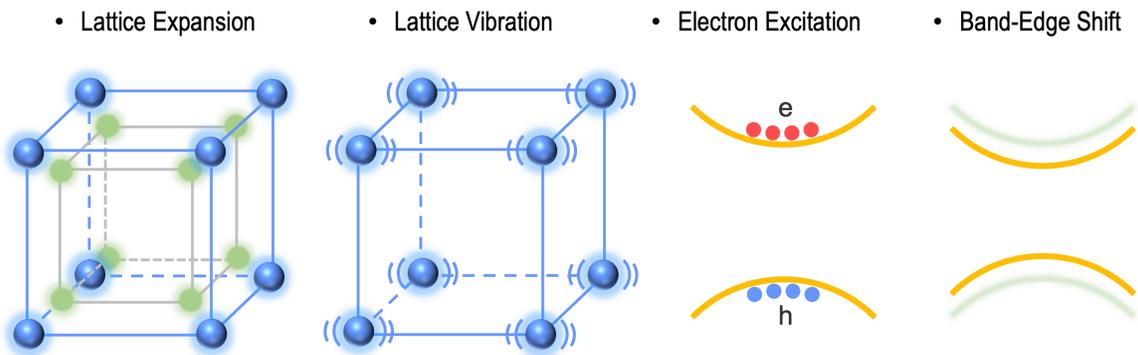

**FIG. 1.** Four types of temperature effects, including lattice expansion, lattice vibration, electron excitation, and band-edge shift, in solids.

In this benchmark study, we employ density functional theory (DFT) calculations to systematically examine all the four types of temperature effects on the hopping distance, defect concentration, attempt frequency, and activation barrier in a consistent manner. By closely examining several distinct diffusion processes, we clarify the impacts of those temperature effects and obtain results with much improved accuracies with respect to the experimental results at high temperatures.

## II. METHODS

Our *ab initio* calculations are carried out based on the spin-polarized DFT as implemented in the Vienna Ab initio Simulation Package.[20] The electron excitation is included by adopting the Fermi-Dirac distribution with an energy smearing consistent with the examined temperature, the lattice expansion is included by using the experimental lattice constants (see section S1 of the supplementary material), and the lattice vibration are calculated with finite displacements and analyzed using the package phonopy.[21] We also obtain the lattice lengths using first-principles calculations and the corresponding diffusivity resembles that based on the experimental lattice length (see section S2 of the supplementary material). For the aluminum system, the exchange-correlation functional is chosen in the form of Perdew-Burke-Ernzerh (PBE).[22] A self-consistent effective Hubbard U term,[23] $U_{eff}$, is added for the *d* electrons in Fe. Our tests show that the Hubbard U term is critical to obtain reasonable hopping barriers (see section S3 of the supplementary material). To accurately describe the electronic properties of 4H-SiC, we employ the Heyd–Scuseria–Ernzerhof hybrid functional (HSE06)[24] for all calculations except the lattice vibration, which is too time-consuming using the hybrid functional and thus is calculated self-consistently with the PBE functional for PBE-relaxed structures. Tests of perfect 4H-SiC bulk show that PBE and HSE06 functionals give very close results (see section S4 of the supplementary material). An energy cutoff of 360, 481, and 450 eV is set to the plane-wave basis sets for the $V_{Al}$, $Fe_{Al}$, and $V_C$ systems, respectively, together with the following projector augmented-wave potentials:[25] Al_GW($3s^23p^1$) for Al, Fe_GW($4s^13d^7$) for Fe, Si_GW($3s^23p^2$) for Si, and C_GW($2s^22p^2$) for C. The *k*-point spacing of $2\pi/60$ and $2\pi/25$ Å$^{-1}$ is used for the Al and the 4H-SiC supercells, which are about 12 and 10 Å in each direction, respectively. Convergence tests of energy cutoff, *k*-point sampling, and supercell size can be found in section S3 of the supplemental material. The hopping barriers are calculated using the climbing nudged elastic band method.[26]

In the *ab initio* molecular dynamics (AIMD) simulation of $V_{Al}$ diffusion, the canonical ensemble is carried out with the experimental lattice constants, and the simulation time varies from 560 to 3640 ps with a time step of 2 fs. The diffusion coefficient is determined by the mean squared displacement (MSD) as defined in Eq. 1,[27]

$$D(T) = \frac{1}{6}\frac{x_D(T)}{x_D^{sim}} \lim_{t \to \infty} \frac{d}{dt} MSD(t). \qquad (1)$$

Here, $x_D^{sim}$ and $x_D(T)$ is the defect concentration in the supercell and that based on Fig. 2(b). The MSD trajectories are analyzed using the Molecular Dynamics Analysis of Neutron Scattering Experiments[28] and the MSD results and diffusion coefficients are provided in section S5 of the supplemental material. The AIMD simulations of 4H-SiC bulk are carried out in the canonical ensemble using the PBE functional with a time step of 1.5 fs for 15 ps; the HSE06 functional is adopted to calculate the temperature dependence of band edges for 11-15 sampled AIMD structures.

## III. RESULTS AND DISCUSSIONS

In this study, we examine three representative dynamical processes at high temperatures, including aluminum vacancy, $V_{Al}$, diffusion in aluminum bulk through vacancy hopping in the range of 643–900 K, diffusion of substitutional defect, $Fe_{Al}$, in aluminum bulk though vacancy-assisted hopping in the range of 500–900 K, and carbon vacancy, $V_C$, diffusion in 4H-SiC bulk through vacancy hopping in the range of 2173–2473 K. These systems are chosen because they represent dynamical processes in non-magnetic metal systems, dilute magnetic systems, and semiconductor systems, and also because they possess sufficient experimental data for an all-aspect verification.

### A. $V_{Al}$ Diffusion in Aluminum Bulk

Our first examined process is $V_{Al}$ diffusion through vacancy hopping in aluminum bulk, which is an ideal system for accurate comparison and fast enough to be verified independently using AIMD. In contrast to the widely-used diffusivity formula shown in Eqs. 2(a-b),[29] we impose the temperature effects to every aspect in the diffusion coefficient of $V_{Al}$, including hopping distance, defect concentration, attempt frequency, and activation barrier, and update the formulas to Eqs. 3(a-b).

$$D(T) = \Gamma f a^2 x_D v_0 e^{-\frac{E_b}{k_B T}} \quad (2a)$$

$$x_D(T) = e^{-\frac{E_f}{k_B T}}, \quad (2b)$$

$$D(T) = \Gamma f a(T)^2 x_D(T) v_0(T) e^{-\frac{E_b(T)}{k_B T}} \quad (3a)$$

$$x_D(T) = e^{-\frac{G_f(T)}{k_B T}}, \quad (3b)$$

where $\Gamma$ and $f$ are geometric factor and correlation factor, which equals 2 and 0.7815 for FCC structure,[30] respectively; $a$ is hopping distance; $x_D$ is fractional concentration of aluminum vacancy; $v_0$ is attempt frequency; $E_f$ and $G_f$ is the internal energy and Gibbs free energy for a dilute aluminum vacancy, respectively; and $E_b$ is the activation barrier of the hopping process.

The attempt frequency was typically treated in one of two ways: a constant in the range of 1–10 THz[31, 32] or a statistical expression initially for molecular systems (Eq. 4a).[33] Here, we extend the formula to solids by averaging over all the *k*-points in the Brillouin zone and include the temperature dependence, as shown in Eq. 4b.

$$v_0 = \frac{\Pi_i \gamma_i^{IS}}{\Pi_j \gamma_j^{TS}}, \tag{4a}$$

$$v_0(T) = \left(\frac{\Pi_{i,k} \gamma_{i,k}^{IS}(T)}{\Pi_{j,k} \gamma_{j,k}^{TS}(T)}\right)^{1/n_k}, \tag{4b}$$

where $n_k$ is the number of sampled *k*-points in the Brillouin zone; $\gamma_{i,k}^{IS}$ and $\gamma_{j,k}^{TS}$ are, respectively, the positive vibration frequency of the initial and transition state for the phonon branch *j* at point *k* in reciprocal space. Given the definition of transition state, the number of $\gamma_{j,k}^{TS}$ is one less than that of $\gamma_{i,k}^{IS}$ at each *k*-point. Note that since Eqs. 4(a-b) already incorporate the entropy of activation barrier, the $E_b(T)$ term in Eq. 3a should solely represent internal energy. However, if $E_b(T)$ is associated with free energy with entropy contribution, the attempt frequency must be expressed as $\frac{k_B T}{h}$.[34, 35]

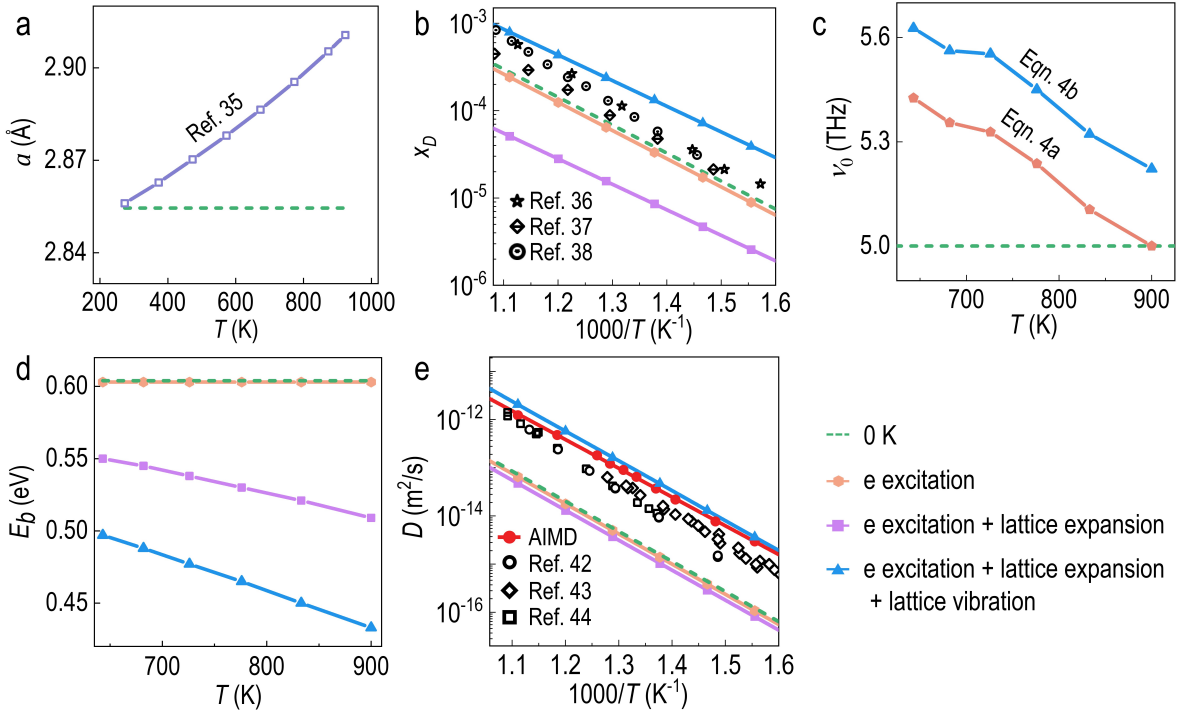

**FIG. 2.** Temperature dependence of the (a) hopping distance, (b) vacancy concentration, (c) attempt frequency, (d) activation barrier, and (e) diffusion coefficient for $V_{Al}$ diffusion in aluminum bulk. In panel (e), an attempt frequency of 5 THz is adopted for the calculations without lattice vibration.

In this work, we consider the thermal expansion by adoption of the experimental lattice constants,[36] which increases by ~2.0% at 900 K relative to that at 300 K [Fig. 2(a)]. Figure 2(b) compares the vacancy concentration predicted at different levels of thermal corrections with the experiments in the range of 643–900 K.[37-39] The electron excitation is found to play a minor role, with the results being 0.85–0.88 times of the 0 K results. Here, "0 K" means none of the lattice expansion, lattice vibration, and electron excitation is included. However, further inclusion of the lattice expansion decreases the vacancy concentration to 0.29–0.21 times of that with electron excitation, and lattice vibration dramatically increases it to 15.25–15.50 times of that with lattice expansion. In total, all the temperature effects induce an increasing vacancy concentration 3.78–2.88 times of that at 0 K.

Figure 2(c) shows that the attempt frequency based on our extended formula of Eq. 4b predicts a slow decrease trend from 5.6 to 5.2 THz in the range of 643–900 K. Relative to the values based on the molecular formula of Eq. 4a, they are around 0.2 THz higher. ~~By contrast, the classical mechanics formula of Eq. 4a incorrectly predicts an increase of attempt frequency from 13.40 to 18.75 THz in the same temperature range.~~

Figure 2(d) compares the temperature effects on activation barrier. The electron excitation is found to induce a negligible decrease of around 1 meV relative to that at 0 K. The lattice expansion decreases the hopping barrier by 0.05–0.10 eV in the range of 643–900 K, and the lattice vibration further decreases it by 0.05–0.08 eV. In total, the lattice expansion and vibration decrease the hopping barrier almost linearly from 0.11 eV at 643 K to 0.17 eV at 900 K, about twice of the thermal energy $k_\text{B}T$. Relative to previous studies without considering the full temperature effects and/or with potentially insufficient computational accuracy,[13, 19, 40, 41] noticeable differences are observed in the formation energy, migration barrier, and attempt frequency, as shown in section S6 of the supplementary material.

Figure 2(e) further compares the temperature dependence of diffusivity among results with different thermal corrections and experimental data.[42-44] To independently examine the accuracy and computational cost of the thermal corrections, we also carried out AIMD simulations in the temperature range of 643–900 K up to 0.5–3.6 ns. Our results show that the electron excitation and lattice expansion induce slight changes relative to that at 0 K, and they are about 1/15–1/19 times of the experimental values. By contrast, the diffusivities with all the thermal corrections are significantly improved and agree reasonably well with both experiments and the AIMD results. However, the AIMD simulations in this example are about $10^3$ times more time-consuming than calculations with all the thermal corrections. Therefore, this example clearly indicates that including the thermal corrections and utilizing the Eqs. 3a-3b and 4b, we can largely address the underestimated diffusivity at 0 K and predict reliable diffusivity at high temperatures.

### B. Fe$_{Al}$ Diffusion in Aluminum Bulk

To evaluate the thermal corrections for more complex systems, we examine the vacancy-assisted diffusion of Fe$_{Al}$ in Al bulk from 500 to 900 K. The diffusion of such dynamical process is often described by the five-frequency model,[45] which involves five frequencies $\omega_0$–$\omega_4$ related to the hoppings of vacancy far away or near a dilute solute. Because $\omega_2/\omega_1$ is close to zero in this specific case, the five-frequency model can be simplified to three frequencies of $\omega_2$–$\omega_4$ (see section S7 of the supplementary material), the thermal effects of which we will focus on. Here, both the hopping distance and vacancy concentration are same as Fig. 2(a-b). Figure 3(a) shows that the attempt frequency varies with processes and exhibits a decreasing trend with temperature, similar to that of aluminum vacancy diffusion in Fig. 2(c). Specifically, the attempt frequency of $\omega_2$–$\omega_4$ decreases from 3.3 to 2.2 THz, from 6.0 to 3.5 THz, and from 9.0 to 8.1 THz, respectively.

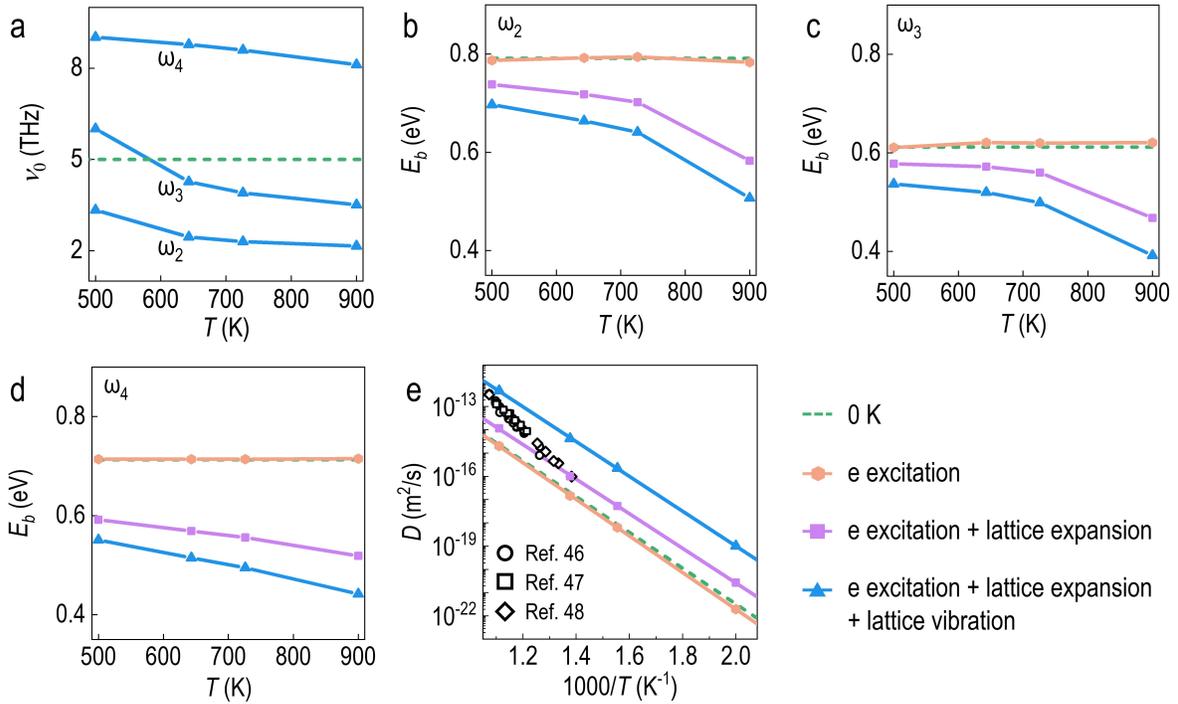

**FIG. 3.** Temperature dependence of (a) attempt frequency and (b–d) activation barriers for processes $\omega_2$–$\omega_4$, and (e) comparison among theoretical diffusion coefficients with different temperature effects and experimental results[46-48] for Fe$_{Al}$ diffusion in aluminum bulk.

For the temperature dependence of hopping barriers, the electron excitation is found to be trivial, with changes less than 10 meV [Fig. 3(b–d)]. By contrast, the lattice expansion decreases the $E_b$ of $\omega_2$–$\omega_4$ from 0.05 to 0.20 eV, from 0.03 to 0.15 eV, and from 0.12 to 0.20 eV in the range of 500–900 K, respectively. The lattice vibration decreases $E_b$ further and the whole temperature effects decrease $E_b$ of $\omega_2$–$\omega_4$ from 0.09 to 0.28 eV, from 0.08 to 0.22 eV, and from 0.16 to 0.27 eV in the range of 500–900 K, respectively. These energy changes are about 1.8–3.6 times of the thermal energy $k_BT$. Compared to previous work with partial temperature effects and without Hubbard U interactions[2], noticeable differences in migration barriers and attempt frequencies are observed (see section S6 of the supplementary material).

Figure 3(e) shows that the total temperature effects increase the diffusion coefficients to 213.6–312.2 times of those at 0 K in the range of 500–900 K. Compared with the experimental results,[46-48] the theoretical predictions with all temperature effects agrees well with the experiments around 900 K, although noticeably higher predictions than the experimental values are observed around 700 K. Two potential reasons could contribute to such discrepancy. First, the error bar of experimental data may be higher at low temperatures because of the significantly slow dynamical processes. Second, the transition metal iron may need an improved description beyond the current Hubbard U method. We compare an effective $U_{eff}$ of 0, an empirical value of 3 eV,[49,50] and a self-consistent value of 2.7 eV for $\omega_2$ at 0 K, and the respective activation barrier is calculated to be 1.33, 0.74, and 0.79 eV (see section S3 of the supplementary material). Although the self-consistent $U_{eff}$ is possibly the optimal choice at hand, we are open to the possibility that further improvement could be made in the future. With improved experimental data, more accurate description of transition metal element, and the high temperature corrections proposed in this work, a better agreement should be reached in the future.

### C. $V_C$ Diffusion in 4H-SiC Bulk

Different from metallic systems, defects in semiconductors or insulators could be charged and experience a transition-state-redox effect during dynamical processes,[1] namely that the transition state possesses a different charge state from that of the initial/final state in certain Fermi level range. Moreover, the band edges of semiconductors and insulators can change dramatically with temperature and consequently impact the Fermi energy and defect concentration. To evaluate the temperature effects in semiconductors, we examine the $V_C$ diffusion through vacancy hopping in 4H-SiC bulk in the same temperature range of 2373–2573 K as previous experiments.[51] The diffusion coefficient of $V_C$ follows Eq. 3a-b, except that the geometric factor $\Gamma$ is 0.75,[12,52] and the correlation factor is 0.7815 in the direction [0001] of a hexagonal system.[53] The formation energy $G_f$ of a defect $D^q$ with charge state $q$ is updated to Eq. 5 based on the original formula,[54]

$$G_f(D^q, E_F) \equiv G_{tot}(D^q, T) + \delta_{FNV}(D^q, T) - G_{tot}(bulk, T) +$$
$$\sum \mu_i(T) + q[E_{VBM}(bulk, T) + E_F(T)], \quad (5)$$

where $G_{tot}(D^q, T)$ and $G_{tot}(bulk, T)$ are the free energies of the system with defect $D^q$ and perfect bulk at temperature $T$, respectively; $\delta_{FNV}$ is the FNV correction with an experimental dielectric constant of $9.76\varepsilon_0$ to reduce image charge interactions under periodic boundary conditions;[55] In the correction, the experimental dielectric constant is used; $\mu_i(T)$ is the chemical potential of element $i$ to balance the composition between the perfect bulk and defective system, and $\mu_C(T)$ of graphite is used in this case; $E_{VBM}(bulk, T)$ and $E_F(T)$ are the valence band maximum (VBM) energy and Fermi energy relative to the VBM energy at temperature $T$, respectively.

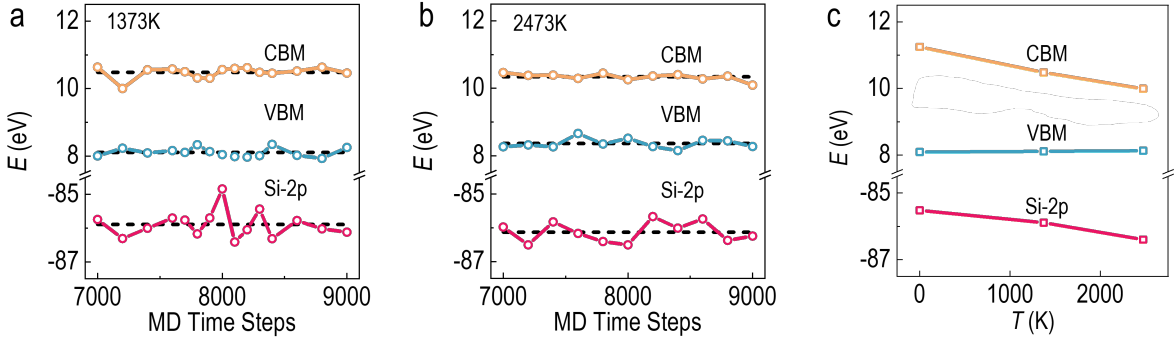

**FIG. 4.** Band edges and Si-2p level for 4H-SiC bulk at sampled time steps for AIMD simulations at (a) 1373 and (b) 2473 K; (c) temperature dependence of band edges and Si-2p for 4H-SiC bulk.

We first examine the temperature dependence of VBM, $E_{VBM}(bulk, T)$. AIMD simulations of 4H-SiC bulk are carried out using the PBE functional with experimental lattice constants for 15 ps and sampled AIMD structures are further calculated using HSE06 functional to obtain the band edges and Si-2p energy level [Fig. 4(a-b)]. Figure 4(c) indicates that the VBM increases slightly with temperature by 0.04 eV at 2473 K relative to that at 0 K, but the CBM decreases dramatically by 1.26 eV. The predicted band gap of 1.9 eV at 2473 K is reasonably consistent with the experimental value of 2.2 eV.[56] The energy level of Si-2p decreases by 0.9 eV from 0 to 2473 K and a linear interpolation in the range of 873-1623 K corresponds to an energy decrease of 0.27 eV, consistent with the value of 0.21 eV according to X-ray photoelectron spectroscopy.[57, 58] Further charge density analyses reveal that the VBM of 4H-SiC is tightly bound to the C atoms, but the CBM is distributed inside the hollow space of the crystal structure (see section S8 of the supplementary material). Such feature is expected to induce easier changes of the CBM by the structural distortion at high temperatures than the VBM.

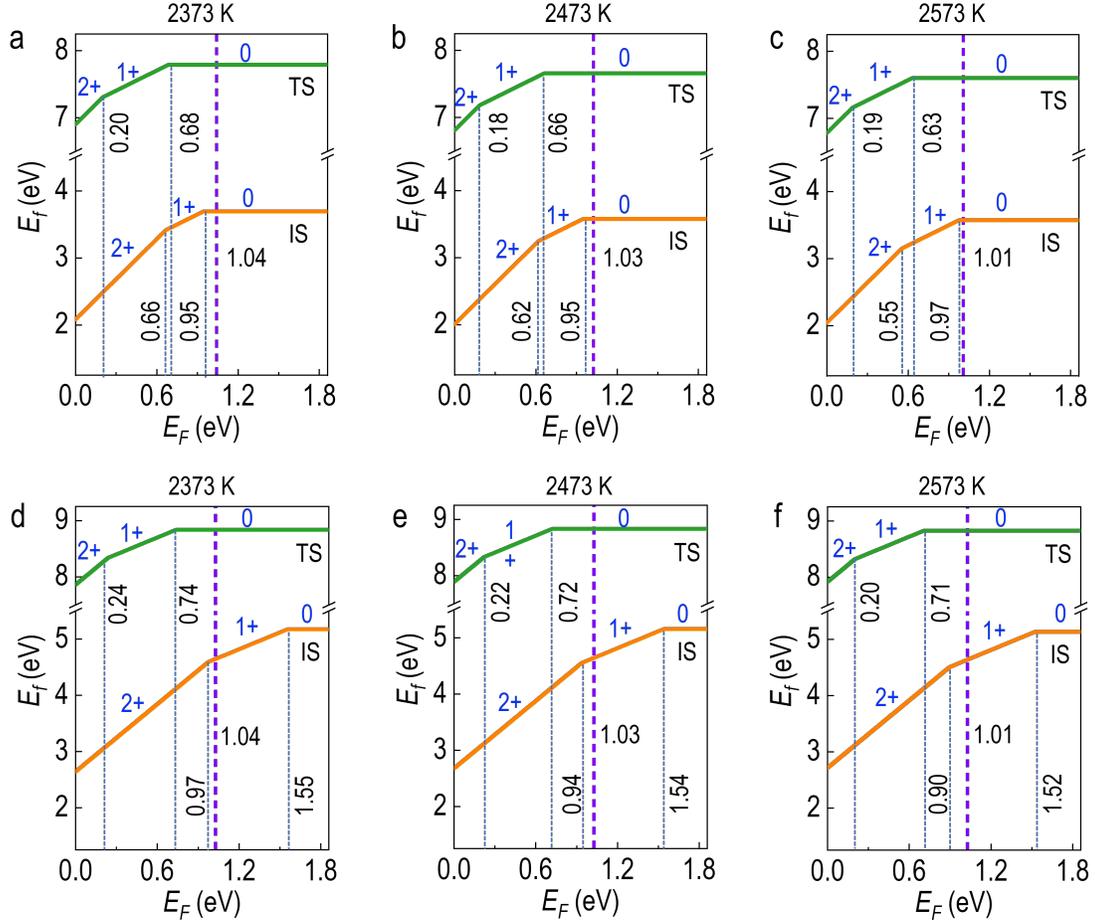

**FIG. 5.** Defect formation energy of the initial state (IS) and transition state (TS) during the $V_C$ hopping in 4H-SiC at (a) 2373, (b) 2473, and (c) 2573 K with all temperature effects and (d-f) corresponding plots without lattice vibration. The blue number beside each segment labels its charge state and the vertical violet dashed line indicates the Fermi level in intrinsic 4H-SiC.

Then we examine the charge state and formation energy of $V_C$ in the initial and transition states, with consideration of the lattice expansion, lattice vibration, electron excitation, and band-edge shift. Figure 5(a) shows that the initial ground state of $V_C$ exhibits 2+, 1+, and 0 charge state in the Fermi level range of 0-0.66, 0.66-0.95, and 0.95-1.86 eV, respectively, at 2373 K. By contrast, the transition state of $V_C$ hopping exhibits 2+, 1+, and 0 charge state in the range of 0-0.20, 0.20-0.68, and 0.68-1.86 eV, respectively. Therefore, the initial and transition state possess different charge states in the Fermi level range of 0.20-0.66 and 0.68-0.95 eV, a phenomenon named as transition state redox [1]. Similar features are also observed at 2473 and 2573 K [Fig. 5(b-c)]. For 4H-SiC bulk, the intrinsic Fermi level is calculated to be 1.04, 1.03, and 1.01 eV at 2373, 2473, and 2573 K, respectively, according to the charge neutrality relationship in Eq. 6.

$$\int_{-\infty}^{E_{VBM}(T)} \frac{DOS[E+E_{VBM}(0)-E_{VBM}(T)]}{1+e^{\frac{(E_F-E)}{k_B T}}} dE = \int_{E_{CBM}(T)}^{\infty} \frac{DOS[E+E_{CBM}(0)-E_{CBM}(T)]}{1+e^{\frac{(E-E_F)}{k_B T}}} dE, \qquad (6)$$

where $E_{VBM}(T)$ and $E_{CBM}(T)$ are the VBM and CBM energy at $T$, respectively; $DOS(E)$ is the density of state at 0 K. Here, a rigid shift of the valence bands and conduction bands obtained at 0 K according to the $E_{VBM}(T)$ and $E_{CBM}(T)$ in Fig. 4(c) is utilized. More rigorous calculations using the average DOS of AIMD structures gives similar result (see section S9 of the supplementary material). According to the Fermi levels and Fig. 5(a-c), the charge state of $V_C$ in intrinsic 4H-SiC are 0 in both the initial and transition states under the three temperatures. The corresponding defect formation energies and hopping barriers of $V_C$ can also be easily obtained. For comparison, the calculations with only the electron excitation, lattice expansion, and band-edge shift predict $V_C$ to be 1+ in its initial state [Fig. 5(d-f)], and those without any temperature effects predict it to be 2+ (see section S10 of the supplementary material). For transition state, its charge state is 0 at all levels of thermal corrections. Therefore, the temperature effects are critical in determining the charge state and formation energy of defects in semiconductors and insulators.

Figure 6 summarizes the temperature dependences of hopping distance, vacancy concentration, attempt frequency, hopping barrier, and diffusion coefficient. Experimental result in Fig. 6(a) shows that the lattice expansion of 4H-SiC in the range of 2373–2573 K is about 1.2% relative to that at 300 K.[59] As for the vacancy concentration, the electron excitation slightly decreases it to 0.66–0.68 times of that at 0 K. The lattice expansion increases the vacancy concentration to 2.09–2.35 times of that with only the electron excitation, and the lattice vibration further increases it by 111.52–113.42 times. In total, all the temperature effects increase the vacancy concentration to 153.25–181.30 times of that at 0 K, significantly more than that in aluminum bulk [Fig. 2(b)].

Surprisingly, we find that the attempt frequency in 4H-SiC reaches 530–641 THz in the range of 2373–2573 K [Fig. 6(c)], about 100 times of the typical values in aluminum bulk [Fig. 2(c) and 3(a)] or about 20 times of the Debye frequency in 4H-SiC. Such high attempt frequency can be ascribed to two reasons. First, the large Young's modulus of 4H-SiC (~450 GPa) relative to that of aluminum bulk (~69 Pa) induces much higher phonon frequencies in the former [Fig. 6(d)]. Consequently, the ratio of positive phonon frequencies between the initial and transition state is greater in 4H-SiC than that in aluminum bulk according to Eq. 4b. Second, the phonon DOS of transition state is slightly red-shifted by 0.33–0.43 THz in 4H-SiC [Fig. 6(e)], while in aluminum bulk, the transition state phonon DOS is blue-shifted by about 0.10 THz [Fig. 6(f)]. These noticeable redshifts also contribute to the high attempt frequency in 4H-SiC.

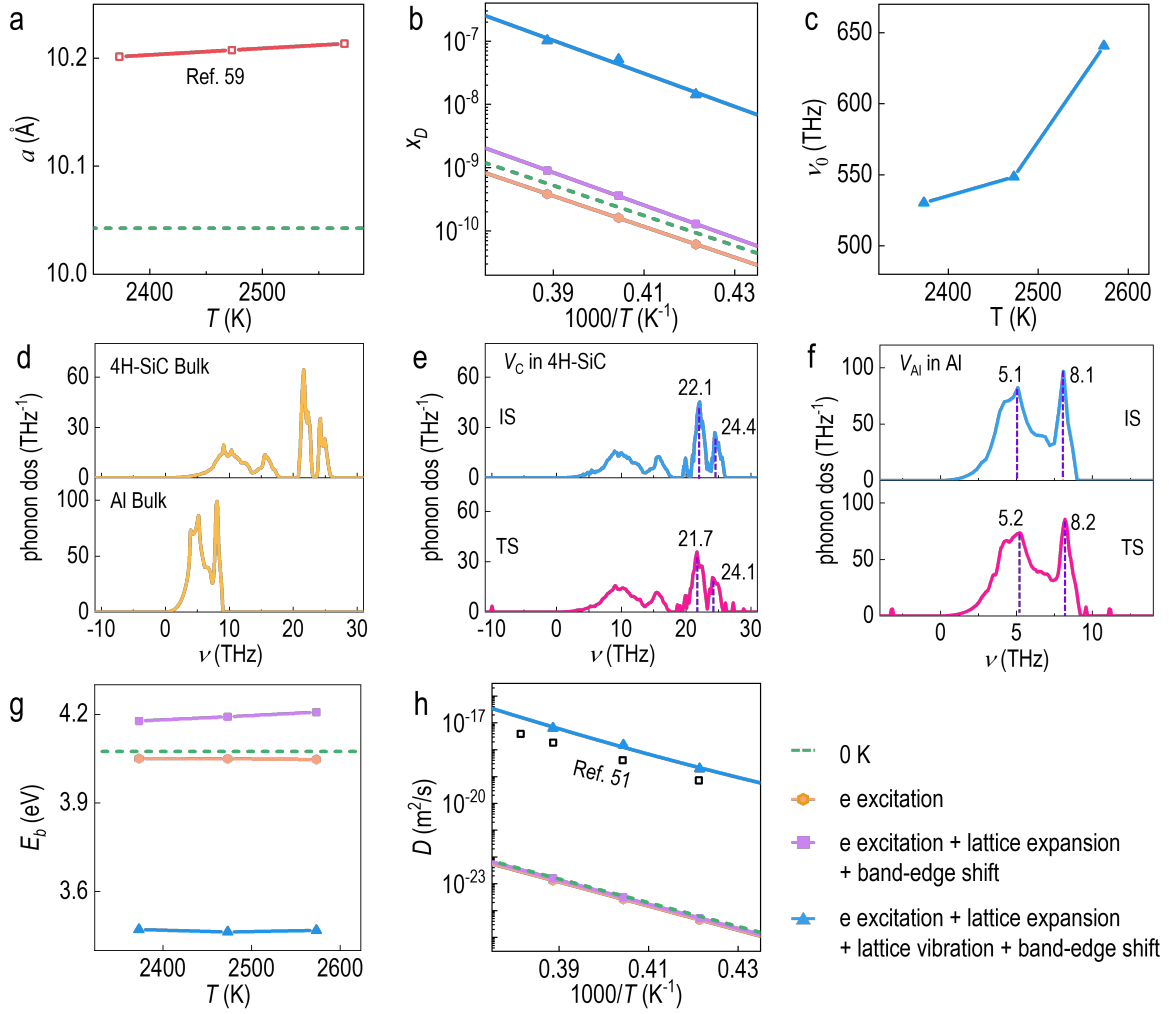

**FIG. 6.** Temperature dependence of (a) hopping distance, (b) vacancy concentration, (c) attempt frequency of $V_C$ diffusion along the [0001] direction of 4H-SiC bulk. Comparison of phonon DOS between (d) 4H-SiC bulk at 2473K and aluminum bulk at 726 K, and phonon DOS of the initial and transition states for (e) $V_C$ hopping in 4H-SiC bulk at 2473K and (f) $V_{Al}$ hopping aluminum bulk at 726 K; (g) hopping barrier and (h) diffusion coefficient of $V_C$ in 4H-SiC bulk. The LO-TO splitting is included in the phonon DOS of 4H-SiC bulk (see the effects of LO-TO splitting in section S11 of the supplementary material).

The hopping barrier of $V_C$ along the [0001] direction of 4H-SiC is found to be 4.07 eV without any temperature effects, and the electron excitation decreases it slightly by 0.02–0.03 eV in the examined temperature range [Fig. 6(g)]. When the lattice expansion is included, the barrier is increased by 0.13–0.16 eV. With the lattice vibration, however, it is decreased by 0.71–0.74 eV. Such reduction of hopping barrier is also observed by comparing with earlier predictions with partial or without any temperature effects (see section S6 of the supplementary material). The overall diffusion coefficient of $V_C$ in [Fig. 6(h)] shows that the result at 0 K with a 5 THz attempt frequency is ~5 orders lower than the experiment.[51] The electron excitation and lattice expansion slightly decrease the diffusion coefficient relative to that at 0 K. By contrast, the lattice vibration increases it by $3.6\times10^5$–$4.0\times10^5$ times. With all the temperature effects, the diffusion coefficient agrees satisfactorily with the experiments.

## IV. CONCLUSION

In summary, we propose a scheme for accurate transition state calculation under high temperatures and apply it to three diffusion processes: $V_{Al}$ diffusion in Al bulk, substitutional defect $Fe_{Al}$ diffusion in Al bulk, and $V_C$ diffusion in 4H-SiC bulk. Our first-principles calculations reveal that electron excitation has a minor influence on diffusion coefficient, and lattice expansion induces a limited improvement on the prediction of diffusion coefficient. The lattice vibration, however, exhibits a remarkable impact on diffusion coefficient. For semiconductors and insulators, band-edge shift can be significant at high temperatures and is critical to determine the charge state, defect formation energy, and activation barrier. Contrary to the conventional assumption, the attempt frequency is found to vary with temperature and can differ significantly between materials. Overall, inclusion of all the temperature effects reduce the defect formation energies and activation barriers; the diffusion coefficients are dramatically increased relative to those at 0 K and significantly improved agreements with experiments are obtained.

## SUPPLEMENTARY MATERIAL

See the supplementary material for the experimental lattice constants of aluminum bulk and 4H-SiC bulk; lattice constants of aluminum bulk and 4H-SiC bulk predicted by first-principles calculations and impacts on dynamics; tests of $k$-point mesh, Hubbard U value, energy cutoff, and supercell size; tests of vibrational free energy of 4H-SiC bulk; AIMD results of $V_{Al}$ diffusion in aluminum bulk; five-frequency model for $Fe_{Al}$ diffusion in aluminum bulk; comparison with theoretical literature; charge density of VBM and CBM in 4H-SiC bulk; DOS and intrinsic Fermi level of 4H-SiC bulk at 2473 K; defect formation energy of $V_C$ in 4H-SiC bulk at 0 K; effects of LO-TO splitting on the phonon DOS, attempt frequency, and diffusivity of $V_C$ in 4H-SiC bulk at 2473 K.

## ACKNOWLEDGEMENT

This work was financially supported by the Guangdong Provincial Key Laboratory of Computational Science and Material Design (Grant No. 2019B030301001), the Introduced Innovative R&D Team of Guangdong (Grant No. 2017ZT07C062), and the Shenzhen Science and Technology Innovation Committee (No. JCYJ20200109141412308). The first-principles calculations were carried out on the Taiyi cluster supported by the Center for Computational Science and Engineering of Southern University of Science and Technology and also on The Major Science and Technology Infrastructure Project of Material Genome Big-science Facilities Platform supported by Municipal Development and Reform Commission of Shenzhen.

## AUTHOR DECLARATIONS

### Conflict of Interest

The authors have no conflicts to disclose.

### Author Contributions

**Chengxuan Ke:** Data curation (lead); Formal analysis (equal); Methodology (supporting); Writing – original draft (lead); Writing – review & editing (supporting). **Chenxi Nie:** Formal analysis (supporting). **Guangfu Luo:** Conceptualization (lead); Formal analysis (equal); Funding acquisition (lead); Methodology (lead); Resources (lead); Supervision (lead); Writing – review & editing (lead).

## DATA AVAILABILITY

The data that support the findings of this study are available from the corresponding author upon reasonable request.

## REFERENCES

1. G. Luo, T. F. Kuech and D. Morgan, "Transition state redox during dynamical processes in semiconductors and insulators," NPG Asia Mater. **10**, 45-51 (2018).
2. H. Wu, T. Mayeshiba and D. Morgan, "High-throughput ab-initio dilute solute diffusion database," Sci Data **3**, 160054 (2016).
3. D. Murali, B. K. Panigrahi, M. C. Valsakumar and C. S. Sundar, "Diffusion of Y and Ti/Zr in bcc iron: a first principles study," J. Nucl. Mater. **419**, 208-212 (2011).
4. M. Krcmar, C. Fu, A. Janotti and R. Reed, "Diffusion rates of 3d transition metal solutes in nickel by first-principles calculations," Acta Mater. **53**, 2369-2376 (2005).
5. N. Sandberg and R. Holmestad, "First-principles calculations of impurity diffusion activation energies in Al," Phys. Rev. B **73**, 014108 (2006).
6. O. Buggenhoudt, T. Schuler, C. C. Fu and J. L. Bechade, "Predicting carbon diffusion in cementite from first principles," Phys. Rev. Mater. **5**, 063401 (2021).
7. J. Li, S. M. Zhang, C. R. Li, Y. G. Zhu, J. A. Boscoboinik, X. Tong, J. T. Sadowski, G. F.

Wang and G. W. Zhou, "Coupling between bulk thermal defects and surface segregation dynamics," Phys. Rev. B **104**, 085408 (2021).

8. C. Li, E. S. Sanli, D. Barragan-Yani, H. Stange, M. D. Heinemann, D. Greiner, W. Sigle, R. Mainz, K. Albe, D. Abou-Ras and P. A. van Aken, "Secondary-Phase-Assisted Grain Boundary Migration in $CuInSe_2$," Phys. Rev. Lett. **124**, 095702 (2020).

9. P. Saidi, P. Changizian, E. Nicholson, H. K. Zhang, Y. Luo, Z. W. Yao, C. V. Singh, M. R. Daymond and L. K. Beland, "Effect of He on the Order-Disorder Transition in $Ni_3Al$ under Irradiation," Phys. Rev. Lett. **124**, 075901 (2020).

10. A. C. P. Jain, P. A. Burr and D. R. Trinkle, "First-principles calculations of solute transport in zirconium: Vacancy-mediated diffusion with metastable states and interstitial diffusion," Phys. Rev. Mater. **3**, 033402 (2019).

11. Y. L. Shi, J. Q. Qi, Y. Han and T. C. Lu, "Anisotropic Diffusion of a Charged Tritium Interstitial in $Li_2TiO_3$ from First-Principles Calculations," Phys. Rev. Appl **10**, 024021 (2018).

12. S. Ganeshan, L. G. Hector and Z. K. Liu, "First-principles study of self-diffusion in hcp Mg and Zn," Comput. Mater. Sci. **50**, 301-307 (2010).

13. K. Carling, G. Wahnstrom, T. R. Mattsson, A. E. Mattsson, N. Sandberg and G. Grimvall, "Vacancies in metals: from first-principles calculations to experimental data," Phys. Rev. Lett. **85**, 3862-3865 (2000).

14. E. Wimmer, W. Wolf, J. Sticht, P. Saxe, C. B. Geller, R. Najafabadi and G. A. Young, "Temperature-dependent diffusion coefficients from ab initio computations: hydrogen, deuterium, and tritium in nickel," Phys. Rev. B **77**, 134305 (2008).

15. L. T. Kong and L. J. Lewis, "Transition state theory of the preexponential factors for self-diffusion on Cu, Ag, and Ni surfaces," Phys. Rev. B **74**, 073412 (2006).

16. K. Sato, S. Takizawa and T. Mohri, "Theoretical calculation of activation free energy for self-diffusion in prototype crystal," Mater Trans **51**, 1521-1525 (2010).

17. F. Legrain, O. I. Malyi and S. Manzhos, "Comparative computational study of the diffusion of Li, Na, and Mg in silicon including the effect of vibrations," Solid State Ion **253**, 157-163 (2013).

18. W. Frank, U. Breier, C. Elsasser and M. Fahnle, "First-Principles Calculations of Absolute Concentrations and Self-Diffusion Constants of Vacancies in Lithium," Phys. Rev. Lett. **77**, 518-521 (1996).

19. B. Grabowski, L. Ismer, T. Hickel and J. Neugebauer, "Ab initio up to the melting point: Anharmonicity and vacancies in aluminum," Phys. Rev. B **79**, 134106 (2009).

20. G. Kresse and J. Furthmuller, "Efficient iterative schemes for ab initio total-energy calculations using a plane-wave basis set," Phys. Rev. B **54**, 11169-11186 (1996).

21. A. Togo and I. Tanaka, "First principles phonon calculations in materials science," Scr. Mater. **108**, 1-5 (2015).

22. J. P. Perdew, K. Burke and M. Ernzerhof, "Generalized gradient approximation made simple," Phys. Rev. Lett. **77**, 3865-3868 (1996).

23. M. Cococcioni and S. de Gironcoli, "Linear response approach to the calculation of the effective interaction parameters in the LDA+U method," Phys. Rev. B **71**, 035105 (2005).

24. J. Heyd, G. E. Scuseria and M. Ernzerhof, "Hybrid functionals based on a screened Coulomb potential," J. Chem. Phys. **118**, 8207-8215 (2003).

25. P. E. Blochl, "Projector Augmented-Wave Method," Phys. Rev. B **50**, 17953-17979 (1994).

26. G. Henkelman, B. P. Uberuaga and H. Jonsson, "A climbing image nudged elastic band method for finding saddle points and minimum energy paths," J. Chem. Phys. **113**, 9901-9904 (2000).

27. M. I. Mendelev and Y. Mishin, "Molecular dynamics study of self-diffusion in bcc Fe," Phys. Rev. B **80**, 144111 (2009).

28. G. Goret, B. Aoun and E. Pellegrini, "MDANSE: an interactive analysis environment for molecular dynamics simulations," J. Chem. Inf. Model. **57**, 1-5 (2017).

29. N. L. Peterson, "Self-Diffusion in Pure Metals," J. Nucl. Mater. **69-7**, 3-37 (1978).

30. G. E. Murch and R. J. Thorn, "Calculation of the diffusion correlation factor by Monte Carlo methods," Philos. Mag. (Abingdon) **39**, 673-677 (1979).

31. X. L. Li, P. Wu, R. J. Yang, D. Yan, S. Chen, S. P. Zhang and N. Chen, "Boron diffusion in bcc-Fe studied by first-principles calculations," Chin. Phys. B **25**, 036601 (2016).

32. T. Maeda and T. Wada, "First-principles study on diffusion of Cd in $CuInSe_2$," MRS Proceedings **1538**, 21-25 (2013).

33. G. H. Vineyard, "Frequency factors and isotope effects in solid state rate processes," J. Phys. Chem. Solids **3**, 121-127 (1957).

34. H. Eyring, "The activated complex in chemical reactions," J. Chem. Phys. **3**, 107-115 (1935).

35. S. Glasstone, K. J. Laidler and H. Eyring, *The theory of rate processes; the kinetics of chemical reactions, viscosity, diffusion and electrochemical phenomena*, 1st ed. (McGraw-Hill Book Company, inc., New York; London, 1941).

36. A. J. C. Wilson, "The thermal expansion of aluminium from 0° to 650 °C," Proc. Phys. Sci. **53**, 235-244 (1941).

37. T. Hehenkamp, "Absolute Vacancy Concentrations in Noble-Metals and Some of Their

Alloys," J. Phys. Chem. Solids **55**, 907-915 (1994).

38. A. Seeger, "Investigation of Point-Defects in Equilibrium Concentrations with Particular Reference to Positron-Annihilation Techniques," J. Phys. F: Met. Phys. **3**, 248-295 (1973).
39. R. O. Simmons and R. W. Balluffi, "Measurements of Equilibrium Vacancy Concentrations in Aluminum," Phys. Rev. **117**, 52-61 (1960).
40. N. Sandberg, B. Magyari-Kope and T. R. Mattsson, "Self-diffusion rates in Al from combined first-principles and model-potential calculations," Phys. Rev. Lett. **89**, 065901 (2002).
41. M. Mantina, Y. Wang, R. Arroyave, L. Q. Chen, Z. K. Liu and C. Wolverton, "First-Principles Calculation of Self-Diffusion Coefficients," Phys. Rev. Lett. **100**, 215901 (2008).
42. M. Beyeler and Y. Adda, "Détermination des volumes d'activation pour la diffusion des atomes dans l'or, le cuivre et l'aluminium," J. Phys. **29**, 345-352 (1968).
43. S. Dais, R. Messer and A. Seeger, "Nuclear-magnetic-resonance study of self-diffusion in aluminium," Mater. Sci. Forum **15-18**, 419-424 (1987).
44. T. S. Lundy and J. F. Murdock, "Diffusion of $Al^{26}$ and $Mn^{54}$ in aluminum," J. Appl. Phys. **33**, 1671-1673 (1962).
45. R. E. Howard and J. R. Manning, "Kinetics of solute-enhanced diffusion in dilute face-centered-cubic alloys," Phys. Rev. **154**, 561 (1967).
46. G. M. Hood, "The diffusion of iron in aluminium," Philos. Mag. **36**, 305-328 (1970).
47. W. B. Alexander and L. M. Slifkin, "Diffusion of solutes in aluminum and dilute aluminum alloys," Phys. Rev. B **1**, 3274 (1970).
48. G. Rummel, T. Zumkley, M. Eggersmann, K. Freitag and H. Mehrer, "Diffusion of implanted 3D-transition elements in aluminium part II: pressure dependence," Int. J. Mater. Res. **86**, 131-140 (1995).
49. P. Mohn, C. Persson, P. Blaha, K. Schwarz, P. Novak and H. Eschrig, "Correlation induced paramagnetic ground state in FeAl," Phys. Rev. Lett. **87**, 196401 (2001).
50. L. Wang, T. Maxisch and G. Ceder, "Oxidation energies of transition metal oxides within the GGA+U framework," Phys. Rev. B **73**, 195107 (2006).
51. M. K. Linnarsson, M. S. Janson, J. Zhang, E. Janzen and B. G. Svensson, "Self-diffusion of $^{12}C$ and $^{13}C$ in intrinsic 4H-SiC," J. Appl. Phys. **95**, 8469-8471 (2004).
52. M. Beyeler and D. Lazarus, "Activation Volume Measurements," Zeitschrift für Naturforschung A **26**, 291-299 (1971).
53. J. Philibert, *Atom movements diffusion and mass transport in solids*. (Editions de Physique, Les Ulis, France, 1991).

54. C. Freysoldt, B. Grabowski, T. Hickel, J. Neugebauer, G. Kresse, A. Janotti and C. G. Van de Walle, "First-principles calculations for point defects in solids," Rev. Mod. Phys. **86**, 253 (2014).
55. C. Freysoldt, J. Neugebauer and C. G. Van de Walle, "Fully ab initio finite-size corrections for charged-defect supercell calculations," Phys. Rev. Lett. **102**, 016402 (2009).
56. M. E. Levinshtein, S. L. Rumyantsev and M. S. Shur, *Properties of advanced semiconductor materials GaN, AlN, InN, BN, SiC, SiGe*. (John Wiley, New York, 2001).
57. S. Diplas, M. Avice, A. Thogersen, J. S. Christensen, U. Grossner, B. G. Svensson, O. Nilsen, H. Fjelivag, S. Hinderc and J. F. Watts, "Interfacial studies of $Al_2O_3$ deposited on 4H-SiC(0001)," Surf Interface Anal **40**, 822-825 (2008).
58. J. Rozen, M. Nagano and H. Tsuchida, "Enhancing interface quality by gate dielectric deposition on a nitrogen-conditioned 4H-SiC surface," J. Mater. Res. **28**, 28-32 (2013).
59. M. Stockmeier, R. Muller, S. A. Sakwe, P. J. Wellmann and A. Magerl, "On the lattice parameters of silicon carbide," J. Appl. Phys. **105**, 033511 (2009).

# Supplemental Material for "High-Temperature Effects for Transition State Calculations in Solids"

Chengxuan Ke[1], Chenxi Nie[1], and Guangfu Luo[1,2*]

[1]Department of Materials Science and Engineering, Southern University of Science and Technology, Shenzhen 518055, China.

[2]Guangdong Provincial Key Laboratory of Computational Science and Material Design, Southern University of Science and Technology, Shenzhen 518055, China.

*E-mail: luogf@sustech.edu.cn

## S1. Experimental lattice constants of aluminum bulk and 4H-SiC bulk

Tables S1 and S2, respectively, list the lattice constants of aluminum bulk and 4H-SiC bulk at different temperatures based on experimental results.[1, 2]

**Table S1.** Lattice constants of aluminum bulk measured at different temperatures.[1]

| $T$ (°C) | 0 | 100 | 200 | 300 | 400 | 500 | 600 | 650 |
|---|---|---|---|---|---|---|---|---|
| $a = b = c$ (Å) | 4.0391 | 4.0486 | 4.0592 | 4.0701 | 4.0820 | 4.0947 | 4.1087 | 4.1162 |

**Table S2.** Lattice constants of 4H-SiC bulk calculated accordingly to the formula obtained experimentally.[2]

| $T$ (°C) | 1900 | 2100 | 2200 | 2300 |
|---|---|---|---|---|
| $a$ (Å) | 9.339 | 9.350 | 9.356 | 9.361 |
| $c$ (Å) | 10.189 | 10.201 | 10.208 | 10.214 |

## S2. Lattice constants of aluminum bulk and 4H-SiC bulk predicted by first-principles calculations and impacts on dynamics

The lattice parameters can also be predicted using first-principles calculations according to the free energy diagram as a function of the lattice constants and temperature. Here, the free energy consists of the internal energy and the vibrational free energy. The PBE and HSE06 functional is used for the Al bulk and 4H-SiC bulk, respectively. As shown in Fig. S1(a-b), the predicted lattice length of Al bulk is about 0.09 Å or 0.8% greater than the experiments in the range of 643-900 K, and the predicted lattice length of 4H-SiC is about 0.07 Å or 0.8% greater than the experiments in the range of 2173-2573 K. Tests of the self-diffusion in Al bulk at 643 K show that the predicted lattice length leads to the hopping distance, attempt frequency, and $\exp[-\frac{G_f(T)+E_b(T)}{k_B T}]$ being 1.01, 0.94, and 0.82 times of the respective value obtained using the experimental lattice length. Overall, the self-diffusion coefficient at 643 K is 0.78 times of the value based on the experimental lattice length, as shown in Fig. S1(c).

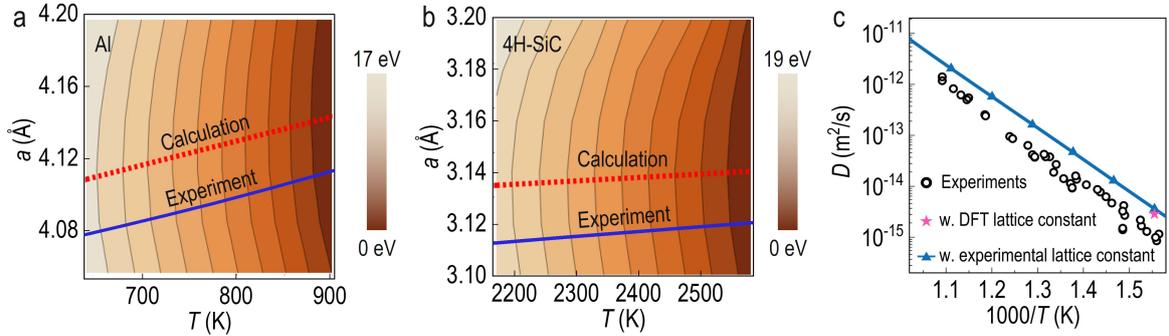

**FIG. S1.** Free energy diagram as a function of lattice length and temperature for (a) aluminum bulk and (b) 4H-SiC bulk; the red dashed line is the predicted lattice length based on the minimum of free energy diagram and the blue line corresponds to the experimental values;[1, 2] (c) comparison of self-diffusion coefficient in aluminum bulk among experiments,[3-5] prediction based on experimental lattice constants, and prediction based on the DFT lattice length in (a).

**S3. Tests of *k*-point mesh, Hubbard U value, energy cutoff, and supercell size**

Figure S2(a) shows the convergence test of *k*-point mesh for the formation energy and hopping barrier of $V_{Al}$ in aluminum bulk at 0 K. According to these tests, a *k*-point mesh of 5 × 5 × 5 is finally used, and the corresponding formation energy and hopping barrier are 0.63 and 0.61 eV, and the corresponding error bars are less than 18 and 3 meV, respectively.

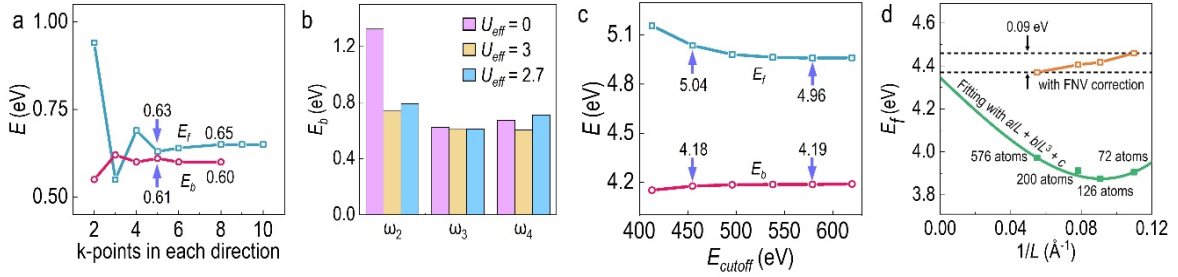

**FIG. S2.** (a) *k*-point test of $V_{Al}$ formation energy and hopping barrier in aluminum bulk at 0 K. (b) Effect of $U_{eff}$ on $E_b$ of $\omega_2$–$\omega_4$ in the diffusion of $Fe_{Al}$ in aluminum bulk at 0 K. (c) Energy cutoff test for the formation energy $E_f$ and hopping barrier $E_b$ of $V_C^{2+}$ in 4H-SiC with HSE06 functional at 0 K. (d) Supercell size tests of $V_C^{2+}$ formation energy in 4H-SiC bulk at the PBE functional level. The Fermi level is set to the middle of band gap and $L$ is defined as $V^{1/3}$, where $V$ is the supercell volume. With the FNV correction, the error of a 72-atom supercell is around 0.09 eV relative to the value at the dilute limit (1/$L$ → 0), which is also extrapolated from the raw data using polynomial.[6]

Figure S2(b) compares the hopping barriers of $\omega_2$–$\omega_4$ at 0 K for $Fe_{Al}$ diffusion in Al bulk using different Hubbard U values. We find that the $\omega_2$ process is particularly sensitive to $U_{eff}$: the activation barrier is 1.33 and 0.74 eV with $U_{eff}$ of 0 and an empirical value of 3 eV,[7,8] respectively. Such large difference highlights the importance of accurate description of 3*d* electrons and probably explains the significant underestimation of the $Fe_{Al}$ diffusion in aluminum bulk in literature.[9] A self-consistent method based on the linear response approach[10] predicts an $U_{eff}$ of 2.7 eV for $Fe_{Al}$ in Al bulk. This approach was designed to comply with a property of the exact density functional: the total energy as a function of electron occupation exhibits a piecewise linearity.[10,11] Therefore, we choose $U_{eff}$ = 2.7 eV for our final calculations, and the corresponding activation barriers are 0.79, 0.61, and 0.71 eV for the $\omega_2$-$\omega_4$ processes, respectively.

Figure S2(c) shows the convergence test of energy cutoff for the formation energy $E_f$ and hopping barrier $E_b$ of $V_C^{2+}$ in 4H-SiC bulk at 0 K. It turns out that an energy cutoff of 450 eV converges the formation energy and hopping barrier at the level of 0.08 and 0.01 eV, respectively, which are supposed to be low.

Figure S2(d) shows the supercell size tests for the $V_C^{2+}$ in 4H-SiC bulk. It is found that, with the FNV correction, the formation energy of a 72-atom supercell is ~0.09 eV or ~2.0% greater than that of the 576-atom supercell. Since the image charge interaction decreases as the charge decrease, the error bars for $V_C^+$ and $V_C^0$ are expected to be even less.

## S4. Tests of vibrational free energy of 4H-SiC bulk

Figure S3 compares the vibrational free energies per formula of 4H-SiC bulk calculated self-consistently with the PBE and HSE06 functional. The PBE and HSE06 results agree well in the temperature range of 0–2500 K and follows a linear relationship. Such agreement supports the reliability of the PBE vibrational results.

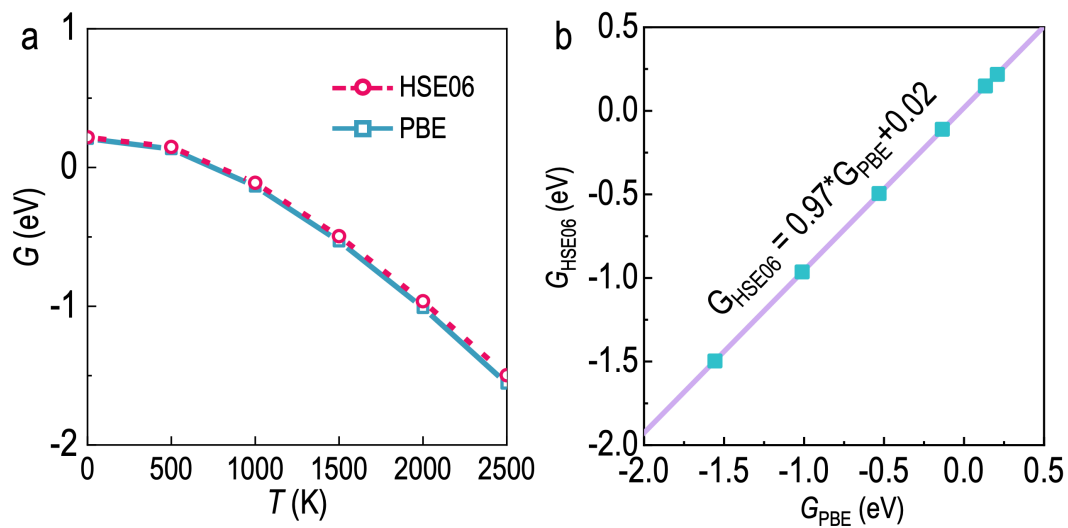

**FIG. S3.** (a) Vibrational free energies per formula of 4H-SiC bulk versus temperature for the PBE and HSE06 results. (b) Linear relationship between the PBE and HSE06 results.

## S5. AIMD results of $V_{Al}$ diffusion in aluminum bulk

Figure S4(a) shows the mean squared displacements (MSD) of $V_{Al}$ diffusion in aluminum bulk obtained at different temperatures. The first $2\times10^4$ time-steps have been discarded to ensure that the system reaches thermal equilibrium, and the nonlinear part of the MSD data at the end of AIMD simulation is not considered for linear fitting because of the insufficient data points. The calculated diffusion coefficient exhibits a perfect linear relationship as shown in Fig. S4(b).

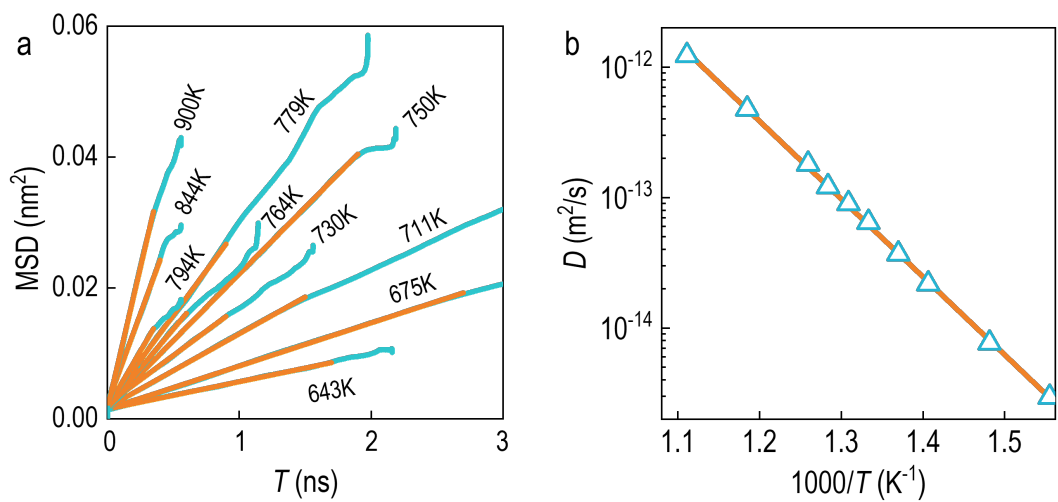

**FIG. S4.** (a) MSD results from 643 to 900 K and (b) temperature dependence of diffusion coefficient for $V_{Al}$ diffusion in aluminum bulk

To clarify if there is any diffusion path different from the vacancy-hopping mechanism utilized in our transition state calculations, we examine the atomic coordinates before and after each hopping, as exemplified in Fig. S5(a-c). It is found that all the hoppings exhibit a hopping distance of 2.75±0.25 Å (Fig. S5d), which is consistent with the neighboring Al-Al distance of 2.91 Å. Further direct visualization of the hopping events also confirms the vacancy-hopping mechanism for the self-diffusion in Al bulk.

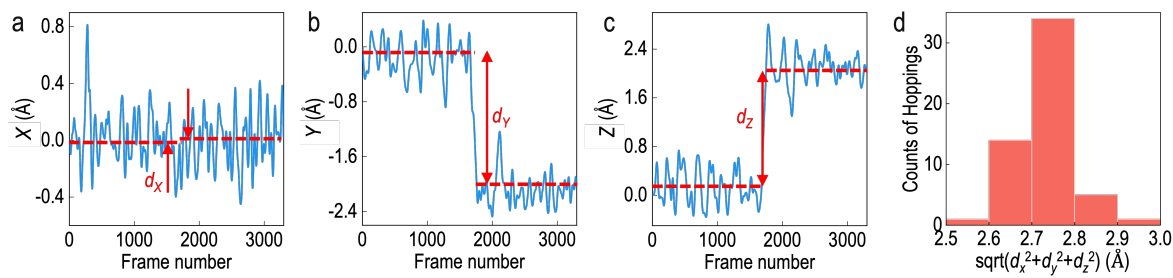

**FIG. S5.** Atomic coordinates of (a) $X$, (b) $Y$, and (c) $Z$ direction of one hopping atom in the AIMD simulation of $V_{Al}$ in Al bulk at 900 K. Dashed line represents the mean value before and after a hopping. (d) Statistics of hopping distance in the 900 K AIMD simulation with 55 hoppings in total.

## S6. Comparison with theoretical literature

**Table S3.** Comparison of defect formation energy, activation enthalpy, attempt frequency, and diffusion coefficient of $V_{Al}$ hopping in Al bulk, vacancy-assisted $Fe_{Al}$ diffusion in Al bulk, and $V_C$ hopping in 4H-SiC between this work and theoretical literature.

| Dynamics | $G_f$ (eV) | $E_b$ (eV) | $v_0$ (THz) | $D$ (m$^2$/s) | Supercell size, energy cutoff, functional/method, temperature effects | Ref. |
|---|---|---|---|---|---|---|
| $V_{Al}$ hopping in Al bulk (643 K ≤ T ≤ 900 K) | 0.63–0.60[i] | — | — | — | 64 atoms, 204 eV, PW91 Lattice vibration | 12 |
| | 0.57–0.57 0.60–0.56[ii] | — | — | — | 108 atoms, 190 eV/462 eV, PBE Lattice expansion Lattice vibration Electron excitation | 13 |
| | 0.62–0.60 | 0.61[i] | 8.4 (Debye frequency) | 5×10$^{-16}$– 5×10$^{-13}$ | 80 atoms, 130 eV, PW91 Lattice vibration | 14 |
| | 0.63–0.61[i] | 0.57[i] | 19.3 | 9×10$^{-16}$– 7×10$^{-13}$ | 32 atoms, 300 eV, PBW Lattice expansion Lattice vibration | 15 |
| | 0.56–0.55 | 0.50–0.43 | 5.63–5.22 | 4×10$^{-15}$– 2×10$^{-12}$ | 108 atoms, 360 eV, PBE Lattice expansion Lattice vibration Electron excitation | This work |
| $Fe_{Al}$ hopping in Al bulk (500 K ≤ T ≤ 900 K) | 0.48 ($V_{Al}$) | 1.43 ($E_{b2}$) 0.62 ($E_{b3}$) 0.65 ($E_{b4}$) | 3.90 ($v_2$) 4.27 ($v_3$) 5.30 ($v_4$) | 4×10$^{-27}$– 9×10$^{-18}$ | 108 atoms, 350 eV, PBE Lattice vibration for $v_0$ | 9 |
| | 0.56-0.55 ($V_{Al}$) | 0.70–0.51 ($E_{b2}$) 0.54–0.39 ($E_{b3}$) 0.55–0.44 ($E_{b4}$) | 3.34–2.16 ($v_2$) 6.01–3.51 ($v_3$) 9.02–8.12 ($v_4$) | 1×10$^{-19}$– 5×10$^{-13}$ | 108 atoms, 481 eV, PBE+U Lattice expansion Lattice vibration Electron excitation | This work |
| $V_C^0$ hopping in 4H-SiC bulk (2373 K ≤ T ≤ 2573 K) | — | 4.70 | 16 (Debye frequency) | — | 240 atoms, DFTB[iii] Lattice vibration | 16 |
| | 5.03 | 3.78 | — | — | 72 atoms, 450 eV, HSE06 | 17 |
| | 3.69–3.57 | 3.47–3.46 | 517.75–628.58 | 2×10$^{-19}$ – 7×10$^{-18}$ | 72 atoms, 450 eV, HSE06 Lattice expansion Lattice vibration Electron excitation Band-edge shift | This work |

[i]A surface correction of 0.15 and 0.05 eV was added manually to the $G_f$ and $E_b$, respectively.
[ii]Anharmonicity was included through molecular dynamics simulations.
[iii]DFTB: density functional based tight binding

## S7. Five-frequency model for Fe$_{Al}$ diffusion in aluminum bulk

Vacancy-assisted solute diffusion in FCC bulk is typically described by the five-vacancy-jump-frequency model,[18] which involves five processes as indicated in Fig. S6 and quantified by Eqs. S1-S4. Specifically, $\omega_0$ is the jump frequency of a vacancy in pure bulk, $\omega_1$ the jump frequency of a vacancy between nearest-neighbor sites of a solute, $\omega_2$ the jump frequency of the solute-vacancy exchange process, $\omega_3$ the jump frequency of a vacancy from nearest-neighbor sites of solute to non-nearest-neighbor sites, and $\omega_4$ the jump frequency of the reverse jump of process 3. For the diffusion of Fe$_{Al}$ in aluminum bulk, because the hopping barrier of process $\omega_1$ and $\omega_2$ is around 0.38 and 0.74 eV, respectively, $\omega_2/\omega_1$ is close to zero ($2.4\times10^{-4}$–$9.8\times10^{-3}$) in our examined temperature of 500–900 K. Therefore, $f_2$ in Eq. S3 can be approximated as 1, and the diffusion coefficient in Eq. S1 can be solely determined by $\omega_2$, $\omega_3$, $\omega_4$, and $G_f$, as expressed in Eq. 5.

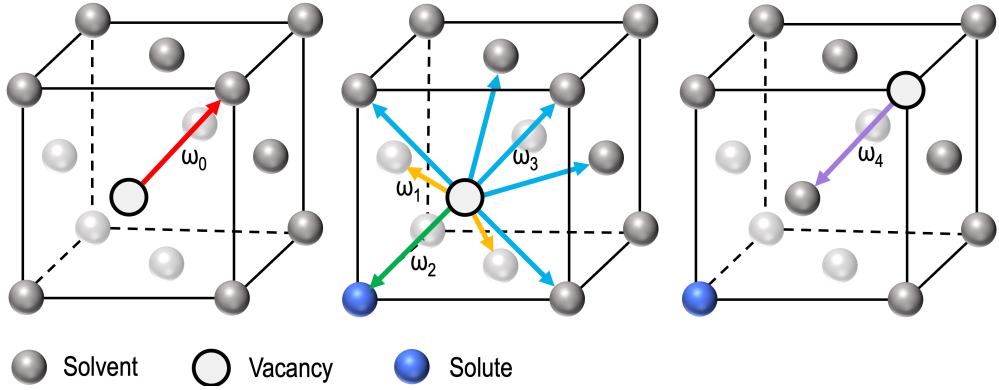

**FIG. S6.** Five-vacancy-jump-frequency model for FCC lattice.

$$D = \frac{f_2\omega_2\omega_4}{f_0\omega_0\omega_3}\left(\Gamma f_0 a^2 e^{\frac{-G_f}{k_BT}}\omega_0\right) \tag{S1}$$

$$\omega_i = v_{0,i} e^{-\frac{E_{bi}(T)}{k_BT}} \tag{S2}$$

$$f_2 = \frac{1+\frac{7}{2}F\left(\frac{\omega_4}{\omega_0}\right)\times\left(\frac{\omega_3}{\omega_1}\right)}{1+\frac{7}{2}F\left(\frac{\omega_4}{\omega_0}\right)\times\left(\frac{\omega_3}{\omega_1}\right)+\left(\frac{\omega_2}{\omega_1}\right)} \tag{S3}$$

$$F(x) = 1 - \frac{10x^4+180.5x^3+927x^2+1341x}{7(2x^4+40.2x^3+254x^2+597x+436)}. \tag{S4}$$

Here, $G_f$ is the formation energy of $V_{Al}$ and $v_{0,i}$ is the $i^{th}$ process' attempt frequency calculated using Eq. 4b in the main text.

$$D = \frac{\omega_2\omega_4}{\omega_3}\left(\Gamma a^2 e^{\frac{-G_f}{k_BT}}\right) \tag{S5}$$

## S8. Charge density of VBM and CBM in 4H-SiC bulk

Figure S7(a-b) shows the charge density of CBM and VBM at 0 K. The CBM majorly locates in the hollow region in the crystal, but VBM binds locally around C atoms. Such spatial distribution does not change radically in the MD simulations at high temperatures, as shown in Fig. S7(c-d). Therefore, the loose (tight) spatial distribution of CBM (VBM) qualitatively explains the fact that the CBM is much more sensitive to the lattice change than VBM as temperature increases.

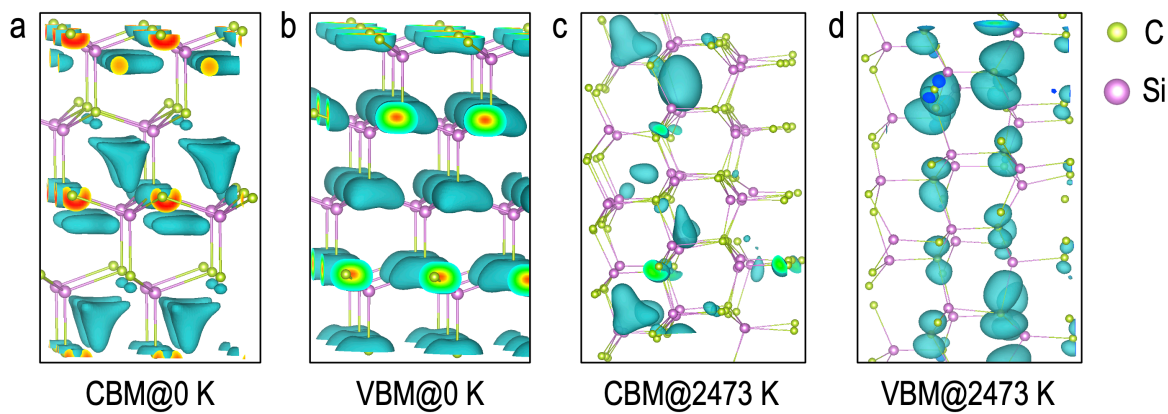

**FIG. S7.** Charge density of (a) CBM and (b) VBM of 4H-SiC bulk at 0 K; examples of charge density of (c) CBM and (d) VBM for 4H-SiC at 2473 K.

## S9. DOS and intrinsic Fermi level of 4H-SiC bulk at 2473 K

Figure S8 compares between the average DOS of 4H-SiC bulk obtained from AIMD structures at 2473 K and the DOS of 0 K with a rigid shift of the conduction bands by -1.31 eV to recover the band gap reduction. Because of the lattice vibration, the average DOS at 2473 K is smoother than that of 0 K. The intrinsic Fermi level calculated based on the average DOS of 2473 K and the shifted DOS of 0 K is 1.029 and 1.026 eV above the VBM, respectively. Given such negligible difference and the high computational cost of average DOS from AIMD structures, the shifted DOS of 0 K is more favorable here.

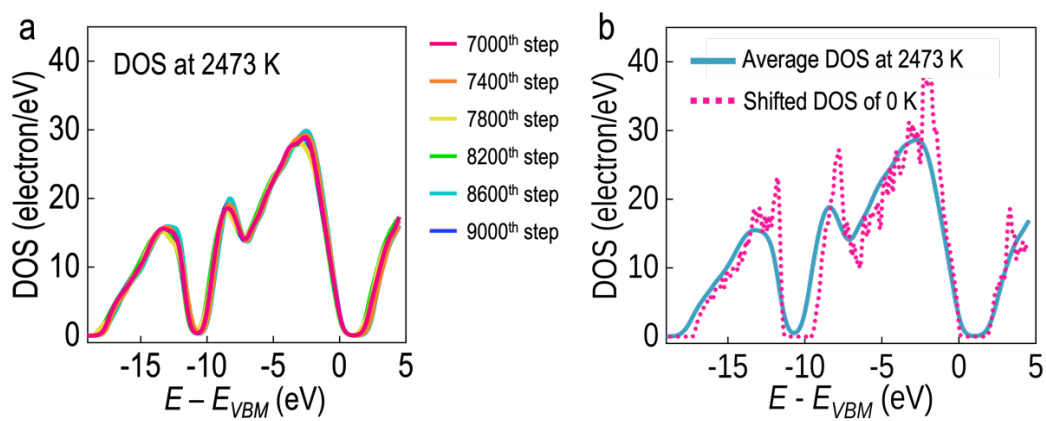

**FIG. S8.** DOS of (a) the 7000th, 7400th, 7800th, 8200th, 8600th, and 9000th step in AIMD simulation at 2473 K, (b) average DOS of the sampled AIMD structures and DOS of 0 K with the conduction bands shifted by -1.31 eV according to Fig. 4c.

## S10. Defect formation energy of $V_C$ in 4H-SiC bulk at 0 K

Figure S9 shows the defect formation energy of $V_C$ in 4H-SiC at 0 K without any temperature effects. The Fermi level is set to be at the mid-gap, where the charge state of $V_C$ is 2+ in the initial state and 0 in the transition state in intrinsic 4H-SiC.

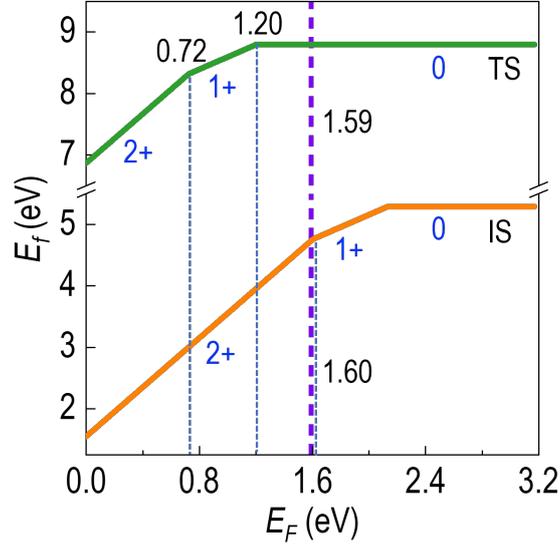

**FIG. S9.** Defect formation energy of the initial state (IS) and transition state (TS) during the $V_C$ hopping in 4H-SiC at 0 K without any temperature effects.

## S11. Effects of LO-TO splitting on the phonon DOS, attempt frequency, and diffusivity of $V_C$ in 4H-SiC bulk at 2473 K

Figure S10(a-b) compares the phonon DOS of $V_C^{2+}$ in 4H-SiC bulk in the initial and transition states at 2473 K with and without the LO-TO splitting. The LO-TO splitting brings in an increase of attempt frequency by ~12 THz or ~2% (Fig. S10c) and an increase of diffusivity by ~2% (Fig. S10d).

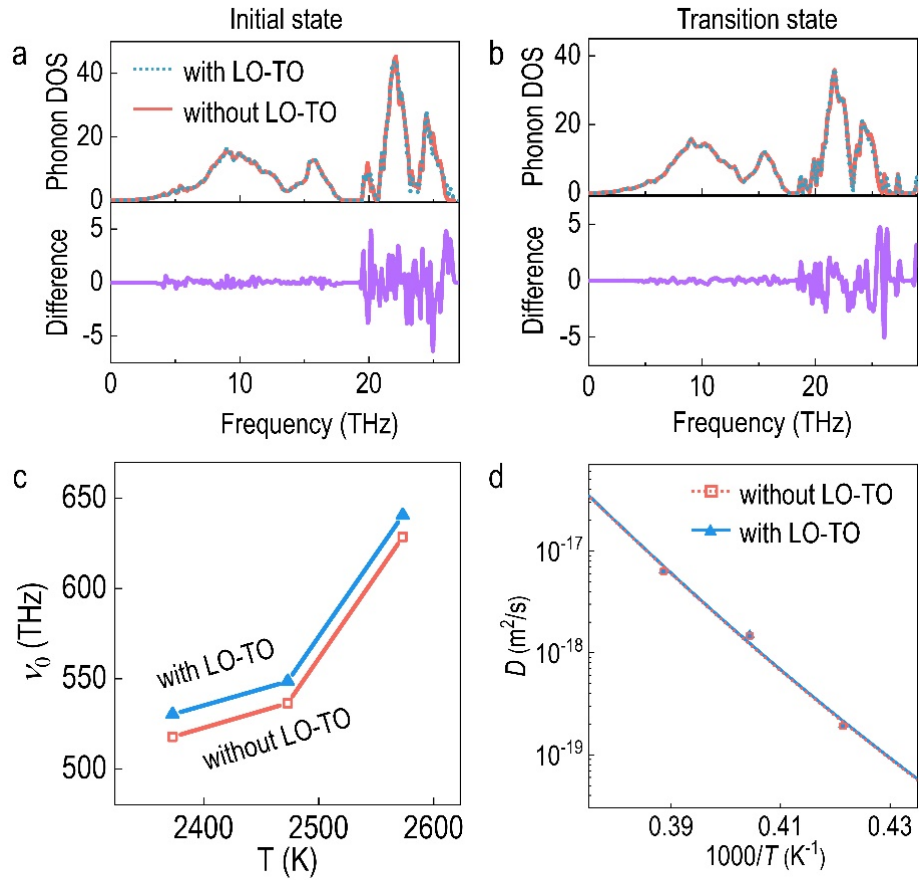

**FIG. S10.** Comparison of phonon DOS of $V_C^{2+}$ in 4H-SiC at 2473 K for the (a) initial state and (b) transition state with and without LO-TO splitting; comparison of the (c) attempt frequency and (d) diffusivity of $V_C^{2+}$ with and without LO-TO splitting.

**References**

1. A. J. C. Wilson, "The thermal expansion of aluminium from 0° to 650 °C," Proc. Phys. Sci. **53**, 235-244 (1941).

2. M. Stockmeier, R. Muller, S. A. Sakwe, P. J. Wellmann and A. Magerl, "On the lattice parameters of silicon carbide," J. Appl. Phys. **105**, 033511 (2009).

3. M. Beyeler and Y. Adda, "Détermination des volumes d'activation pour la diffusion des atomes dans l'or, le cuivre et l'aluminium," J. Phys. **29**, 345-352 (1968).

4. S. Dais, R. Messer and A. Seeger, "Nuclear-magnetic-resonance study of self-diffusion in aluminium," Mater. Sci. Forum **15-18**, 419-424 (1987).

5. T. S. Lundy and J. F. Murdock, "Diffusion of $Al^{26}$ and $Mn^{54}$ in aluminum," J. Appl. Phys. **33**, 1671-1673 (1962).

6. G. Makov and M. C. Payne, "Periodic Boundary-Conditions in Ab-Initio Calculations," Phys. Rev. B **51**, 4014-4022 (1995).

7. P. Mohn, C. Persson, P. Blaha, K. Schwarz, P. Novak and H. Eschrig, "Correlation induced paramagnetic ground state in FeAl," Phys. Rev. Lett. **87**, 196401 (2001).

8. L. Wang, T. Maxisch and G. Ceder, "Oxidation energies of transition metal oxides within the GGA+U framework," Phys. Rev. B **73**, 195107 (2006).

9. H. Wu, T. Mayeshiba and D. Morgan, "High-throughput ab-initio dilute solute diffusion database," Sci Data **3**, 160054 (2016).

10. M. Cococcioni and S. de Gironcoli, "Linear response approach to the calculation of the effective interaction parameters in the LDA+U method," Phys. Rev. B **71**, 035105 (2005).

11. S. Falletta and A. Pasquarello, "Hubbard U through polaronic defect states," npj Comput. Mater. **8**, 263 (2022).

12. K. Carling, G. Wahnstrom, T. R. Mattsson, A. E. Mattsson, N. Sandberg and G. Grimvall, "Vacancies in metals: from first-principles calculations to experimental data," Phys. Rev. Lett. **85**, 3862-3865 (2000).

13. B. Grabowski, L. Ismer, T. Hickel and J. Neugebauer, "Ab initio up to the melting point: Anharmonicity and vacancies in aluminum," Phys. Rev. B **79**, 134106 (2009).

14. N. Sandberg, B. Magyari-Kope and T. R. Mattsson, "Self-diffusion rates in Al from combined first-principles and model-potential calculations," Phys. Rev. Lett. **89**, 065901 (2002).

15. M. Mantina, Y. Wang, R. Arroyave, L. Q. Chen, Z. K. Liu and C. Wolverton, "First-Principles Calculation of Self-Diffusion Coefficients," Phys. Rev. Lett. **100**, 215901 (2008).

16. E. Rauls, T. Frauenheim, A. Gali and P. Deak, "Theoretical study of vacancy diffusion and vacancy-assisted clustering of antisites in SiC," Phys. Rev. B **68**, 155208 (2003).

17. G. Luo, T. F. Kuech and D. Morgan, "Transition state redox during dynamical processes in semiconductors and insulators," NPG Asia Mater. **10**, 45-51 (2018).

18. R. E. Howard and J. R. Manning, "Kinetics of solute-enhanced diffusion in dilute face-centered-cubic alloys," Phys. Rev. **154**, 561 (1967).